\documentclass{emulateapj}

\newcommand{\figwidth}{1}

\shorttitle{M33 modelling and physical implications}
\shortauthors{Hague and Wilkinson}

\begin{document}

\title{The degeneracy of M33 mass modelling and its physical implications}
\author{P. R. Hague}
\affil{University of Leicester}
\email{peter.hague@le.ac.uk}
\and
\author{M. I. Wilkinson}
\affil{University of Leicester}
\email{miw6@le.ac.uk}
\keywords{dark matter - galaxies: individual (M33) - galaxies: kinematics and dynamics - galaxies: spiral - galaxies: structure - Local Group}

\begin{abstract}
The Local Group galaxy M33 exhibits a regular spiral structure and is close enough to permit high resolution analysis of its kinematics, making it an ideal candidate for rotation curve studies of its inner regions. Previous studies have claimed the galaxy has a dark matter halo with an NFW profile, based on statistical comparisons with a small number of other profiles. We apply a Bayesian method from our previous paper to place the dark matter density profile in the context of a continuous, and more general, parameter space. For a wide range of initial assumptions we find that models with inner log slope $\gamma_{\rm in}<0.9$ are strongly excluded by the kinematics of the galaxy unless the mass-to-light ratio of the stellar components in the $3.6\mu$m band satisfies $\Upsilon_{3.6}\geq2$. Such a high $\Upsilon_{3.6}$ is inconsistent with current modelling of the stellar population of M33. This suggests that M33 is a galaxy whose dark matter halo has not been significantly modified by feedback. We discuss possible explanations of this result, including ram pressure stripping during earlier interactions with M31.
\end{abstract}

\section{Introduction}

Cosmological models of the formation of dark matter haloes predict cusped density profiles \citep{Dubinski:1991ca, Navarro:1996ce}, which do not appear to match the dark matter density profiles inferred from observations of rotation curves of disk galaxies \citep{gentile2004}. 

Decompositions suggesting uniform central density haloes \citep{flores1994, moore1994} led \cite{Burkert:1995jr} to propose a universal, cored profile. Rotation curves were originally measured with a slit along the principal axis of the galaxy, but most current measurements use a tilted ring method to extract rotation curves from velocity fields \citep{Begeman:1989wv}. Using this method \cite{gentile2004} found that cored haloes were preferred to both $\Lambda$CDM haloes and MOND (MOdified Newtonian Dynamics) for a sample of five galaxies. Observations of galaxies from THINGS \citep[The HI Nearby Galaxy Survey;][]{THINGS} have provided improved observational constraints on the rotation curves (and thus density profiles) of nearby galaxies, as explored in \cite{deblok2008} and \cite{Hague:2013bz} (hereafter HW13). These improved velocity data permit more precise constraints on halo density profiles than were possible in previous papers that addressed the cusp-core problem. 

Hydrodynamics simulations have been used to attempt to reconcile dark matter-only $\Lambda$CDM simulations with observations. \cite{Governato:2010ed} found that feedback from supernovae is able to flatten the inner density profile of isolated dwarf galaxies and produce a rotation curve comparable to that observed in the dwarf galaxy DDO 39. In contrast, \cite{Parry:2011fd} found that satellites of a Milky Way-like simulated galaxy, generated using a hydrodynamical simulation that was able to reproduce the observed population and kinematics of the Milky Way system, did not have their dark matter haloes significantly altered by baryonic activity, although they note that they are unable to resolve the innermost parts of the profile due to the force resolution of their simulations. \cite{DiCintio:2013em} found a relation between maximum rotation velocity $v_{\rm rot}$ and the inner log slope of the dark matter profile in 31 simulated galaxies, with cores being seen in smaller ($v_{\rm rot} \sim 50{\rm kms}^{-1}$) galaxies and profiles approaching NFW in larger ($v_{\rm rot} \sim 150{\rm kms}^{-1}$) galaxies.

In this paper we apply the Bayesian method presented in HW13 to the galaxy M33, which previous work has shown can be fitted over the entire radial range of HI data by a single power law $\rho \propto r^{-1.3}$, compatible with the NFW profile \citep{Corbelli:2000jg}. Later work in \cite{Corbelli:2003kn} added molecular gas to the mass model, resulting in an inner density profile $\rho \propto r^{-1.5}$ being excluded. We also examine the more recent claim of \cite{Seigar:2011gm} that the NFW profile itself best represents the dark matter halo of M33. This result is at odds with some observational claims \citep[e.g.][]{deblok2008} that smaller galaxies are best described with cored halo profiles. In \cite{Hague:2014gm} (hereafter HW14) we found, using our Bayesian method, a broad range of inner log slopes in a subset of the THINGS galaxies.

\begin{figure}
  \includegraphics[width=\linewidth]{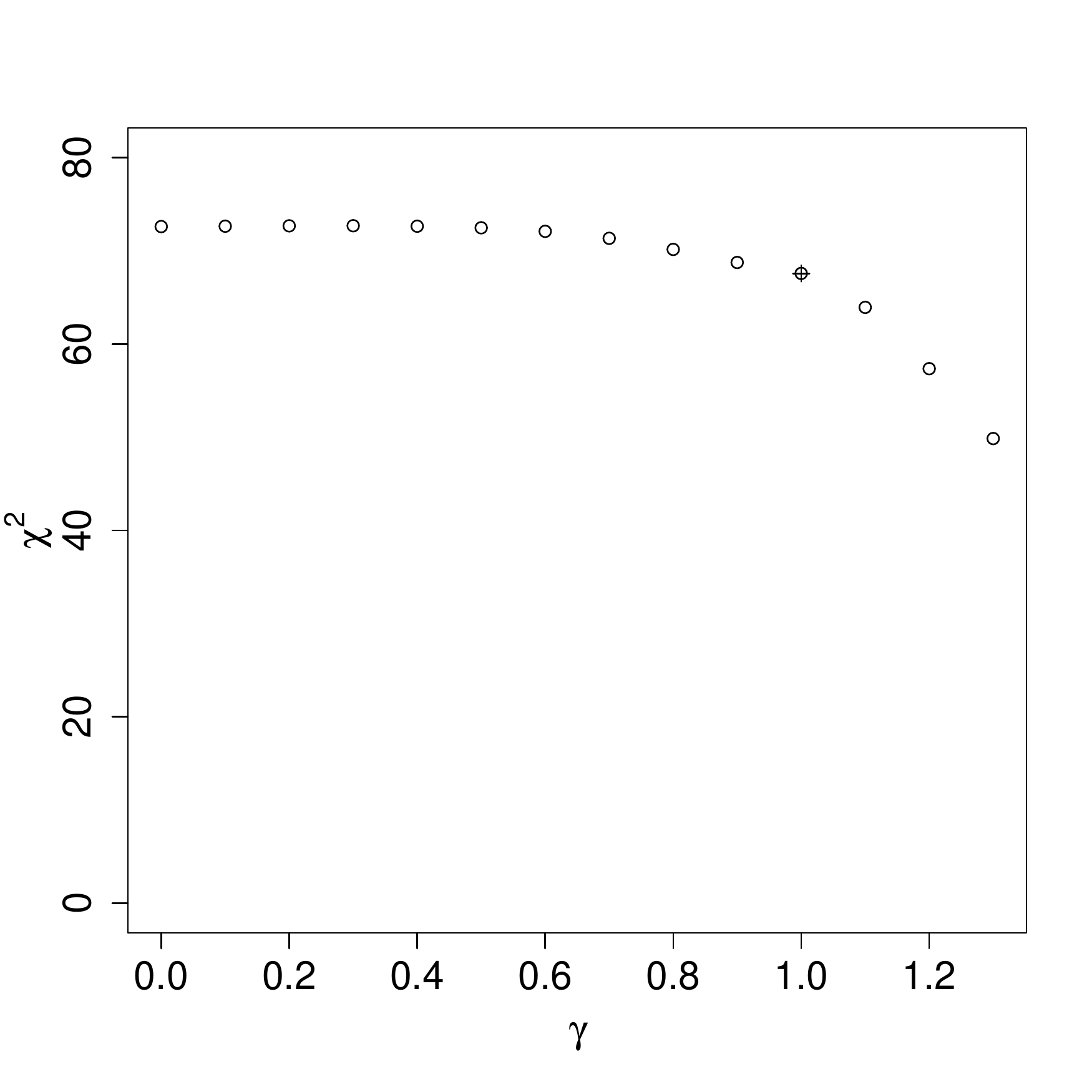}
  \caption{Variability of best fit $\chi^2$, using the rotation curve data presented in Section \ref{datasec}, with inner log slope for a range of modified NFW profiles $\rho \propto (r/r_{\rm s})^\gamma (1+r/r_{\rm s})^{3-\gamma}$, where $\gamma$ is the inner log slope. Free mass-to-light ratios are allowed for both stellar components. The unmodified NFW halo is the marked point.}
  \label{varnfw}
\end{figure}

Previous work on rotation curves \citep[e.g.][]{Chemin:2011iw,gentile2004,Seigar:2011gm} has taken the reduced $\chi^2$ values of mass models to be an accurate representation of the quality of the fit, and further have inferred support for the particular properties of their dark halo models from these values. This is problematic for three main reasons; first the degrees of freedom cannot be trivially inferred from examination of the profile, as it is not clear that all the parameters impact the fit independently at all points in parameter space (see Section \ref{discusssec}). Secondly, even using non-reduced $\chi^2$, the errors that occur in rotation curves derived using the tilted ring model are not Gaussian, and thus the $\chi^2$ statistic is not strictly valid in this case. Thirdly, there are many halo density profiles that vary considerably in their essential qualities that produce comparably good fits measured by $\chi^2$, as shown in Figure \ref{varnfw} where modifying the inner log slope of an NFW profile and then finding the best fit with a free mass-to-light ratio gives good $\chi^2$ for profiles that span the range between cusps and cores. In this context, we present an alternative approach based on MCMC that attempts to overcome these issues, and uses $\chi^2$ as a local estimate of the relative likelihood of nearby models rather than a rigorous global goodness of fit.

In Section \ref{datasec} we present the data we use in this analysis. In Section \ref{methodssec} we describe how we reproduce the baryonic mass modelling and rotation curve of M33, and the MCMC technique we use with this model. In Section \ref{resultssec} we analyse the output of the MCMC chains and in Section \ref{discusssec} we discuss our result in the context of previous papers and the current paradigm of galaxy formation. 

\section{M33 Data}
\label{datasec}

We use the rotation curve and gas surface density from \cite{Corbelli:2003kn}. This gas model includes both neutral atomic and molecular gas. The rotation curve is derived from HI velocity cubes of the galaxy using a tilted ring model with 11 free rings. Our stellar luminosity data are taken from \cite{Seigar:2011gm}, which divides the stellar component into a centrally concentrated component (referred to as a bulge in that paper) and a more extended component. 

A more extended rotation curve is shown in \cite{Corbelli:2014wf}, but these data do not provide higher spatial resolution and primarily introduce new circular speed bins in the outer ($>$16kpc) part of the galaxy. As we are focusing on the profile of the inner halo, and the impact baryons have on it, these data are not relevant here. Also, as we explain in \S~\ref{dmmodels}, we specifically use a dark matter density profile that allows for independent fitting at large and small radii, which therefore does not impose a prior relation between the slope of the inner profile and the outer profile.

\section{Modelling of M33}
\label{methodssec}

We decompose the rotation curve of M33 into four components: two stellar disks, a gas disk and a dark matter halo. The circular velocity contribution of each component is added in quadrature to produce a proposed rotation curve, to be compared with observations.  

\subsection{Baryonic Mass Models}
\label{massmodels}

The \cite{Seigar:2011gm} model consists of a gas component taken from \cite{Corbelli:2003kn}, along with two stellar components. These latter components are distinguished photometrically, rather than by velocity structure. The more extended component is assumed to be exponential, whilst the more centrally concentrated component is taken to be a S\'{e}rsic profile (although in this case the best fit was found to be $n=1$, making it equivalent to an exponential profile). Table \ref{masstable} shows the values used to generate the two components. Values for the inner component appear to differ from those quoted in \cite{Seigar:2011gm} as we have converted them from those of a  S\'{e}rsic profile to the equivalents for an exponential disk. We explore models with a freely varying baryonic mass-to-light ratio, $\Upsilon_{3.6}$, the solar masses per solar luminosity in the Spitzer $3.6\mu m$ band, and with a number of fixed mass-to-light ratios taken from previous work or derived from stellar mass modelling. We do not include a mass-to-light gradient in the disk as the estimated gradient in \cite{Seigar:2011gm} ($-0.014{\rm kpc}^{-1}$) gives rise to a change in the total predicted velocity which is less than 0.75 of the observational error bars at all radii. However we investigate the potential impact of varying mass-to-light ratio with radius in models D1 and D2.

Our first model for the stellar mass allows the mass-to-light ratio of the stellar disks to vary freely over a large range. We consider a second, fixed stellar model using the stellar population mass modelling of \cite{Oh:2008ce}, along with the $J-K$ values for M33 taken from the 2MASS Large Galaxy Atlas \citep{Jarrett:2003jz}. These values are based on integrated magnitudes measured within a 20 mag arcsec$^{-2}$ isophote, which in M33 corresponds to a radius $r\approx6{\rm kpc}$. For M33, $[J-K] = 0.891$ which gives a mass-to-light ratio $\Upsilon_{3.6} = 0.67$. This is more consistent with current estimates of mass-to-light ratios of similar nearby galaxies \citep[e.g. ][]{Meidt:2014bg} than the value from \cite{Seigar:2011gm} of $\Upsilon_{3.6} = 1.25$.

We have been provided with the radial surface density of neutral atomic gas and molecular gas used in \cite{Corbelli:2003kn} by the authors. We processed this using the {\tt ROTMOD} task in {\tt GIPSY}\footnote{http://www.astro.rug.nl/$\sim$gipsy/}, which employs the method described in \cite{Casertano:1983uo} to generate a rotation curve contribution. We use a ${\rm sech}^2$ vertical density law, and the value for $z_{\rm gas}=0.5$kpc given in \cite{Corbelli:2000jg}, who note that an infinitesimally thin disk yields an identical result. Other density laws do not produce a sufficiently large difference to impact the analysis, and the gas rotation curve is consistent with that shown in Figure 5 of \cite{Corbelli:2003kn}. The gas contribution used in \cite{Seigar:2011gm} is that of \cite{Corbelli:2000jg}, which does not include molecular gas. However, as shown in \cite{Corbelli:2003kn}, the molecular gas mass is 10\% of the atomic gas and so the result from \cite{Seigar:2011gm} is suitable for a first order comparison.

\subsection{Dark Matter Models}
\label{dmmodels}

To model the dark matter halo we used an $\alpha-\beta-\gamma$ profile, but as in HW13 transformed to remove the degeneracy between $\rho_{\rm s}$ and $r_{\rm s}$

\begin{equation}
\rho(r)= \frac{ \widetilde\Sigma_{\rm max}}{G} \frac{v_{\rm max}^2}{({r \over r_{\rm s}})^\gamma (1+({r \over r_{\rm s}})^{1/\alpha})^{\alpha ( \beta - \gamma)}} 
\end{equation}

where $r_{\rm s}$ is the scale radius, $v_{\rm max}$ is the peak velocity of the dark matter rotation curve, $\alpha, \beta$ and $\gamma$ are shaping parameters, and $\widetilde\Sigma_{\rm max}$ is the normalised surface density at $r_{\rm max} \equiv r(v_{\rm max})$, given by

\begin{equation}
\widetilde\Sigma_{\rm max} = { \rho_{\rm s} r_{\rm max} \over M(r_{\rm max})}
\end{equation}

which means that the parameterisation replaces $\rho_{\rm s}$ with $v_{\rm max}$ but retains the same number of parameters, since $\widetilde\Sigma_{\rm max}$ is fixed at each point in parameter space.

Contrary to the statements in \cite{Adams:2014hj}, it does not matter that the parameters of this halo profile are still degenerate to some extent as we focus on physical properties of each halo model, within the data range, rather than parameters such as $\gamma$. Our approach has the advantage that we do not impose as strong a prior link as all previous authors between the log slope of the halo at small and large radii (due to the larger parameter space) and we do not extrapolate beyond the data range. We have confirmed the utility of this approach through extensive testing in HW13.

\begin{table}
\caption{Parameters used to generate the mass models}
\begin{tabular}{lll}
\hline
Parameter & Definition & Value \\ \hline
$h_{\rm 1}$ (kpc) & Inner stellar disk scale length & 0.235 \\
$M_{\rm 1}$ ($M_\odot$) & Inner stellar disk mass & $6.07\times10^8$ \\
$h_{\rm 2}$ (kpc) & Outer stellar disk scale length & 1.7 \\
$M_{\rm 2}$ ($M_\odot$) & Outer stellar disk mass & $3.81\times10^9$ \\
$z_{\rm gas}$ (kpc) & Gas disk scale height & 0.5 \\
$M_{\rm gas}$ ($M_\odot$) & Gas disk mass & $3\times10^9$ \\
\hline
\end{tabular}
\tablecomments{The two stellar components are modelled as exponential disks using parameters from \cite{Seigar:2011gm}. For definiteness, we use their mass-to-light ratio $\Upsilon_{3.6}=1.25$ here (although can be a free parameter or takes different values in our models, see Table \ref{modelstable}), and the gas component is modelled using radially binned surface density data provided by \cite{Corbelli:2003kn}.}
\label{masstable}
\end{table}

\subsection{MCMC analysis}

We use a Bayesian Markov Chain Monte Carlo (MCMC) method to explore the parameter space of our models. This method produces a non-normalised probability distribution, which can be argued to be normalised if there are no physically credible models outside the parameterisation. The method is described, along with the extensive testing we have done on simulated data, in HW13. Starting at a random position in the parameter space defined by [$\alpha, \beta, \gamma, v_{\rm max}, r_{\rm s}, \Upsilon_{3.6}$], each MCMC chain moves through the space using a Metropolis-Hastings algorithm \citep{HASTINGS:W09WSbSu} which chooses a new model based on a Gaussian step from the existing one, and then moves there if the new model shows a higher likelihood, or with a probability equal to the ratio of the new likelihood divided by the current one if the new likelihood is lower. This results in the chain seeking, and spending most time in, likelihood peaks, but also enables it to move out of peaks to explore other parts of the parameter space. The posterior probability distributions of the parameters are then calculated from the density of models in the parameter space by multiplication with the prior probability. For explicit parameters (e.g. $\alpha$, $\beta$, $\gamma$ etc.) we assume a flat prior to allow the data the greatest freedom to constrain the models. For derived parameters (e.g. $\gamma_{\rm in}$) this leads to an implicit prior which we calculate numerically (see \S 3.1 of HW13).

We use $\chi^2$ to calculated a likelihood value at each chosen point in parameter space, but the validity of our approach only depends on the relative values of $\chi^2$ for nearby models being a reasonable proxy for relative likelihood. There are many halo profiles that produce good $\chi^2$ values for this rotation curve, and the strength of the MCMC method is that it allows us to differentiate between these models and determine what, if any, actual constraint exists. 

Following the method in HW13 we use the publicly available {\tt CosmoMC} code \citep{lewis2002} to implement our MCMC chains. We ran 8 chains in parallel, with a total of $\sim4\times10^7$ models. We have shown in HW14 that this method can be applied to galaxies spanning a wide range of mass and surface brightness.

We present the results for eight runs, shown in Table \ref{modelstable}. For the A runs we allowed a free mass-to-light ratio, $\Upsilon_{3.6}$, with a range [0.1, 5] to generously cover possible stellar contributions from no disk contribution through to a super-maximal disk. For the B runs we model the baryonic components as in \cite{Seigar:2011gm}; for the C runs we use the mass-to-light ratio calculated above (Section \ref{massmodels}), and the D runs use two independent values for $\Upsilon_{3.6}$ for the inner and outer stellar components, using the same ranges as the A runs. 

The parameter space is mirrored around $\gamma=0$, using a range [-2,2] for model selection but taking the absolute value in the range [0,2] for likelihood testing, so that potentially viable cored profiles are not located at the boundary of the parameter space (see HW13 for details). 

\begin{table}
\caption{Parameters of the MCMC runs}
\begin{tabular}{llll}
\hline
Run & $\Upsilon_{3.6}$ & Radial Bins & Number of Models \\ \hline
A1 & [0.1, 5] & 27 & 34969550 \\
B1 & 1.25 & 27 & 39147303 \\
C1 & 0.67 & 27 & 39285713 \\
D1 & [0.1, 5] & 27 & 33891283 \\
\hline
A2 & [0.1, 5] & 25 & 34169361 \\
B2 & 1.25 & 25 & 39634524 \\
C2 & 0.67 & 25 & 39455690 \\
D2 & [0.1, 5] & 25 & 31798101 \\
\hline
\end{tabular}
\tablecomments{The number of radial bins in models in the second set of models is reduced by ignoring the feature at the outermost part of the rotation curve in \cite{Corbelli:2000jg}. The priors for $\Upsilon_{3.6}$ are (1) a freely varying $\Upsilon_{3.6}$ in the range [0.1, 5] (models A, D); (2) the value from \cite{Seigar:2008hl}, who used a central mass-to-light ratio in the Spitzer $3.6\mu m$ band of $\Upsilon_{3.6}=1.25\pm0.10$ (model B); (3) a value for $\Upsilon_{3.6}$ derived in this paper with assumptions from \cite{Oh:2008ce}, as described in the text (model C). }
\label{modelstable}
\end{table}

We have discarded 10,000 models from the beginning of each chain to allow for burn-in, which we find is sufficient for all the chains to move to areas of high likelihood ($\chi^2_{\rm red} < 2.5$). However, after this point some chains explore secondary peaks before finding the main peak. These secondary peaks are a genuine part of the distribution, as can be verified by the fact that chains sometimes leave the main peak to explore them for an extended period. Our use of MCMC in our analysis thus gives us a more complete picture of the multi-modal probability distribution. However, we note that the high $\Upsilon_{3.6}$ tail of the distribution is very weak, and only partially resolved by our chains. Thus, a comparison of likelihood values between these peaks is not meaningful. 

\subsection{Convergence of MCMC Chains}

To check that all 8 chains are converged on the same distribution, we calculated the ratio of the variance of the means of the chains $\sigma(\hat{x})$, to the mean of the variances of the chains $\hat{\sigma}(x)$, for each parameter $x$. This number is not meaningful for $\gamma$, as the distribution is bimodal, but the highest value for other parameters was 0.17 for $\beta$ indicating good convergence. Inspection of the distribution for each chain showed that in some cases the chain had found only the main peak shown in Figure \ref{gammainml} whereas in other cases the chain only found the left hand side of the tail. As some chains managed to integrate the entire distribution as shown in Figure \ref{gammainml}, it is reasonable to assume that given enough time any chain will converge on the same result. Combining the distributions may produce an incorrect relative model density between the density of the main peak and the tail, but our analysis does not make use of these values. The aim here is to include the broadest possible range of alternate hypothesis in order to demonstrate that the areas avoided by the MCMC are truly excluded. Calculating the precise likelihood of each peak explored is not required for this purpose.

\section{Results}
\label{resultssec}

Our main finding is that models of M33 exhibit a well defined degeneracy between stellar mass and halo inner slope. Furthermore, we are able exclude models with both slopes shallower than $\gamma_{\rm in}<0.9$ and stellar mass to light ratios in the $3.6\mu {\rm m}$ band $\Upsilon_{3.6}<2$, for a generous range of priors. Runs A1, A2, D1 and D2 all show the degeneracy, and runs B1, B2, C2 and C3 find values for fixed values of $\Upsilon_{3.6}$ that are consistent with the distribution found with less restrictive mass-to-light priors. We use A1 as our primary example here, and the other runs to demonstrate the insensitivity of the degeneracy to different priors.

\subsection{Individual Profiles}

\begin{figure}
  \includegraphics[width=\linewidth]{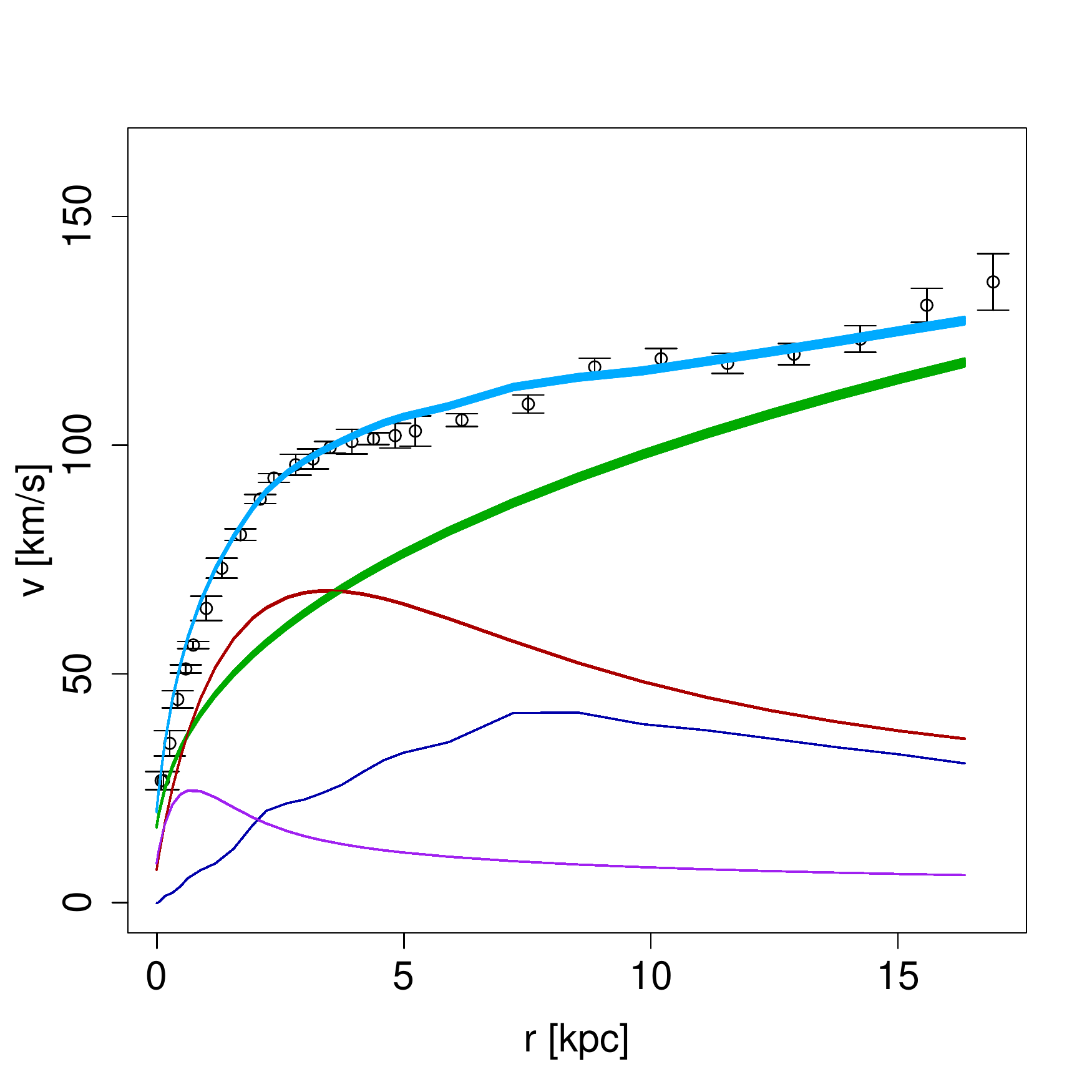}
  \caption{Rotation curve of M33. The dark blue curve is the gas (atomic and molecular) contribution, the purple curve is the inner stellar component, the red curve is the outer stellar component, the green line is a proposed dark matter halo (taken from the most occupied bin from the parameter space of the A1 run) and the light blue line is the expected rotation curve. Observed data from \cite{Corbelli:2003kn} are in black. The mass-to-light ratio of the stellar components has been found by fitting the rotation curve rather than by modelling the stellar population in this case.}
  \label{curve1}
\end{figure}

\begin{figure}
  \includegraphics[width=\linewidth]{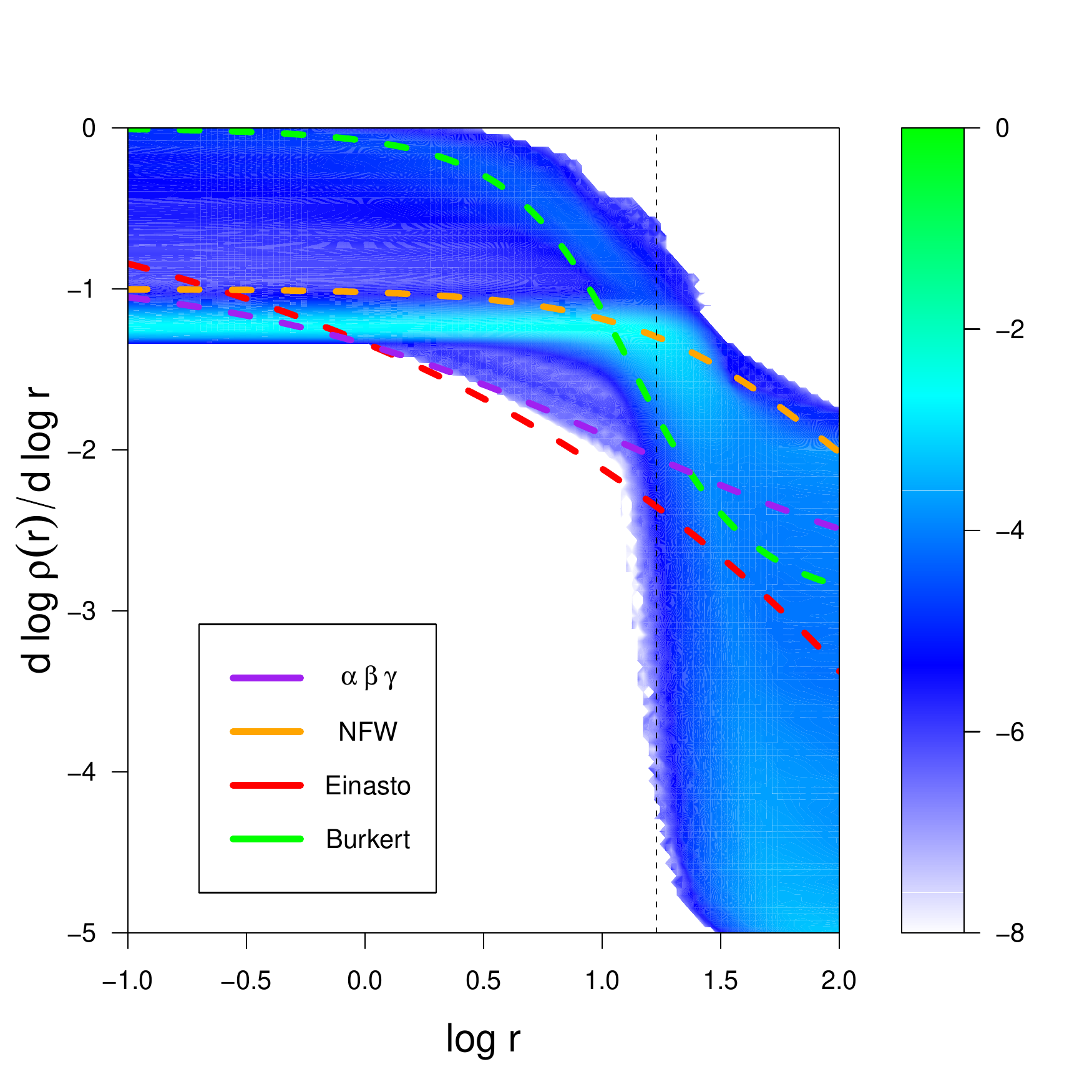}
  \caption{Density plot of all the models produced by A1, in $d log \rho/d log r$ space. Overlaid are fits of commonly used profiles to the M33 data. The vertical dashed line marks
  the outer edge of the observed rotation curve data. }
  \label{overplot}
\end{figure}

A rotation curve from the most populated bin of the A1 distribution is shown in Figure \ref{curve1}, with parameters $(\alpha, \beta, \gamma, r_{\rm s}, v_{\rm max}, \Upsilon_{3.6}) = (0.36, 3.85, 1.22, 32.6{\rm kpc}, 135.4{\rm kms}^{-1}, 1.53)$. A density plot of all the profiles produced in this run is shown in Figure \ref{overplot}, showing the best fits of other commonly used density profiles. The highest density of models occurs in a band centred on a single power-law profile with log-slope $\sim-1.25$, in agreement with the findings of \cite{Corbelli:2000jg}. We note that the sharp lower boundary of the distribution between $\log r =-1$ and $\log r \approx 0.5$ corresponds to the limit of a maximal halo.

\begin{figure}
  \includegraphics[width=\linewidth]{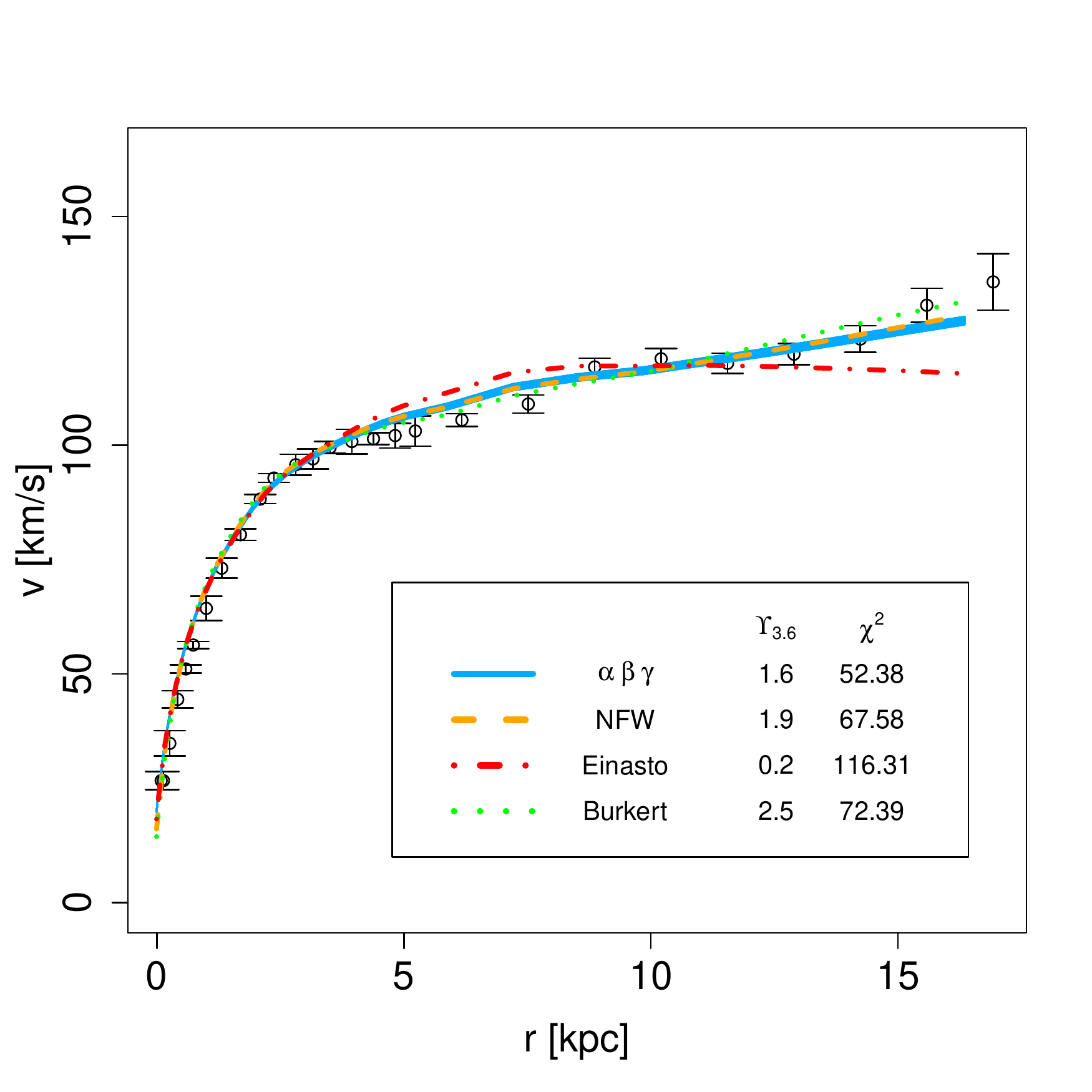}
  \caption{Best models of the M33 rotation curve where the blue band is the best fitting MCMC bin, the green dashed line is the best Burkert halo $\chi^2$ fit, the red dashed line the best Einasto halo fit, and the orange dashed line the best NFW profile fit. In each case the rotation curve was fitted using a free mass-to-light ratio $\Upsilon_{3.6}$. The quoted $\chi^2$ value for the NFW profile is not the same as that found by \cite{Seigar:2011gm} as we use slightly different rotation curve data (see Section \ref{datasec}). The similarity of the $\chi^2$ value for the NFW halo and the Burkert halo (separation of $< 1$ in terms of reduced $\chi^2$) illustrates that this fitting statistic cannot be used to clearly differentiate between cusps and cores in this application and that a more sophisticated technique such as MCMC is needed.}
  \label{allcurves}
\end{figure}

Figure \ref{allcurves} shows the halo of the most favoured part of the parameter space compared to other commonly used halo profiles, which have been fitted by minimising $\chi^2_{\rm red}$, using a free $\Upsilon_{3.6}$ parameter. All the halos are able to capture the data in the inner part of the galaxy reasonably well whilst not fitting the outer data points well. This does not preclude a low value of $\chi^2_{\rm red}$, as the poor fit at large radii can be compensated for by a tight fit at low radii, which is the case in these curves. This further underscores the danger of applying a statistic such as $\chi^2_{\rm red}$ to these data.

\subsection{Log Slope Degeneracy}

We measure the inner log slope of each model at the innermost data point, $\gamma_{\rm in}$, rather than relying on the parameter $\gamma$, so that our measurement of the slope is not an extrapolation outside the data range. 

\begin{figure}
  \includegraphics[width=\linewidth]{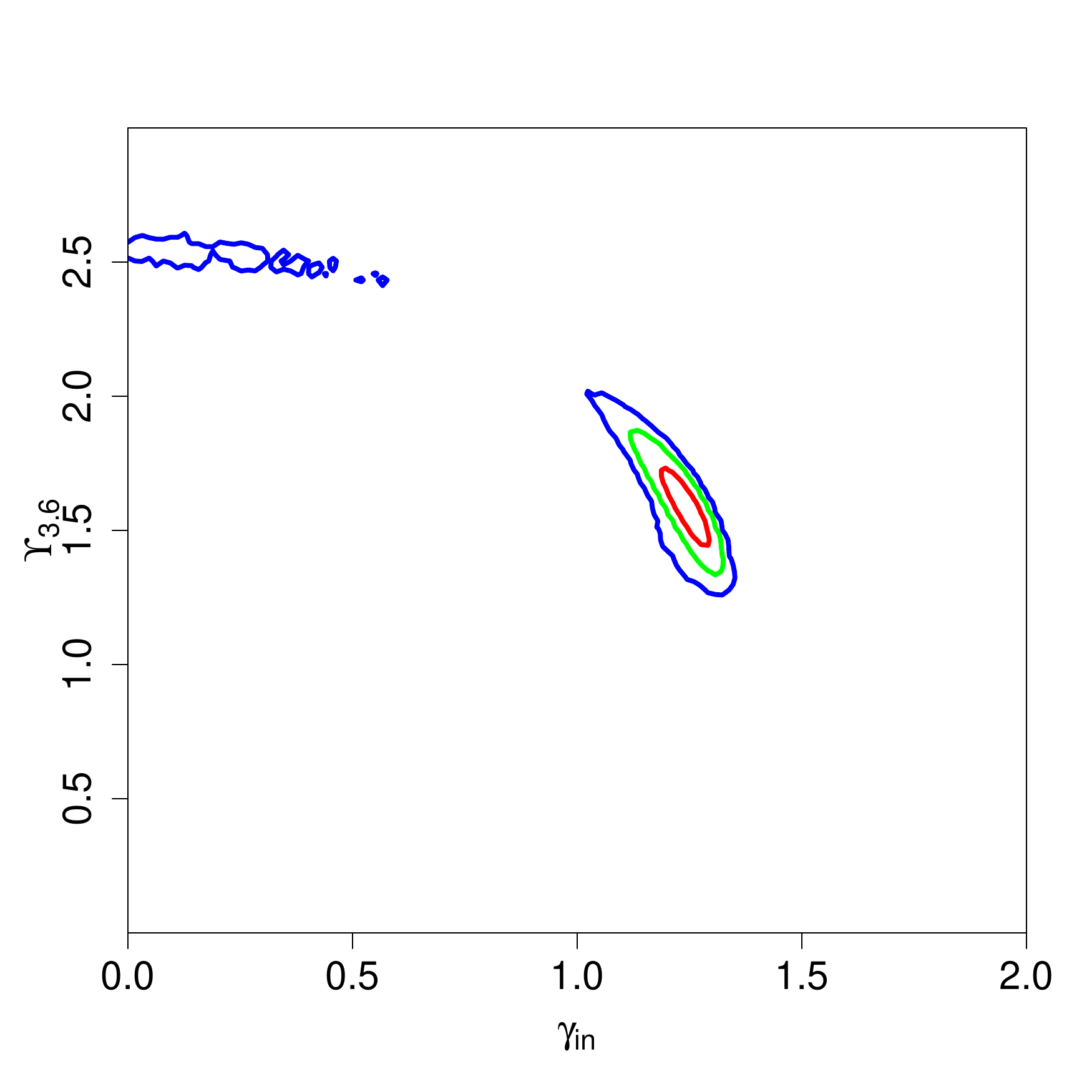}
  \caption{Contour plot of the mass-to-light $\Upsilon_{3.6}$ versus inner log slope $\gamma_{\rm in}$. The red contour contains 68\% of all the models in the MCMC chains, the green contour contains 95\% and the blue contour contains 99\%.}
  \label{gammainml}
\end{figure}

For A1, we found $\gamma_{\rm in}$ to be degenerate with the mass-to-light ratio as shown in Figure \ref{gammainml}. The distribution is binned on a $128\times128$ grid and then contours placed that enclose 68\%, 95\%, and 99\% of all models. Our analysis favours models with steep inner cusps and high mass-to-light ratios, with a tail in the distribution moving towards flat haloes with even higher values of $\Upsilon_{3.6}$. The two areas are connected by a bridge of models which is not shown here as the density is below the $3\sigma$ level. The exact combination of stellar disk mass and dark matter halo favoured in \cite{Seigar:2011gm} ($\Upsilon_{3.6}=1.25$ and $\gamma_{\rm in} \simeq 1$) is disfavoured at over $3\sigma$ when using the free prior here. However, this does not mean that it cannot fit the rotation curve, or that it is not the most favoured result given a more constraining prior on $\Upsilon_{3.6}$. The question we address here is not whether the NFW halo can fit the data (it can, as has been established in the work of \cite{Seigar:2011gm} and \cite{Corbelli:2014wf} and confirmed here). Rather, we are asking whether other models provide better fits, and thus what can actually be inferred from the fact that a particular profile does fit the data.

At the low $\gamma_{\rm in}$ end of the plot is the maximal disk case, where the baryonic component of the galaxy contributes almost all of the rotation curve. Note that despite being flat, this region is not flush with the upper boundary of the $\Upsilon_{3.6}$ range - higher values are excluded by the rotation curve data themselves.

\begin{figure*}
  \includegraphics[width=\linewidth]{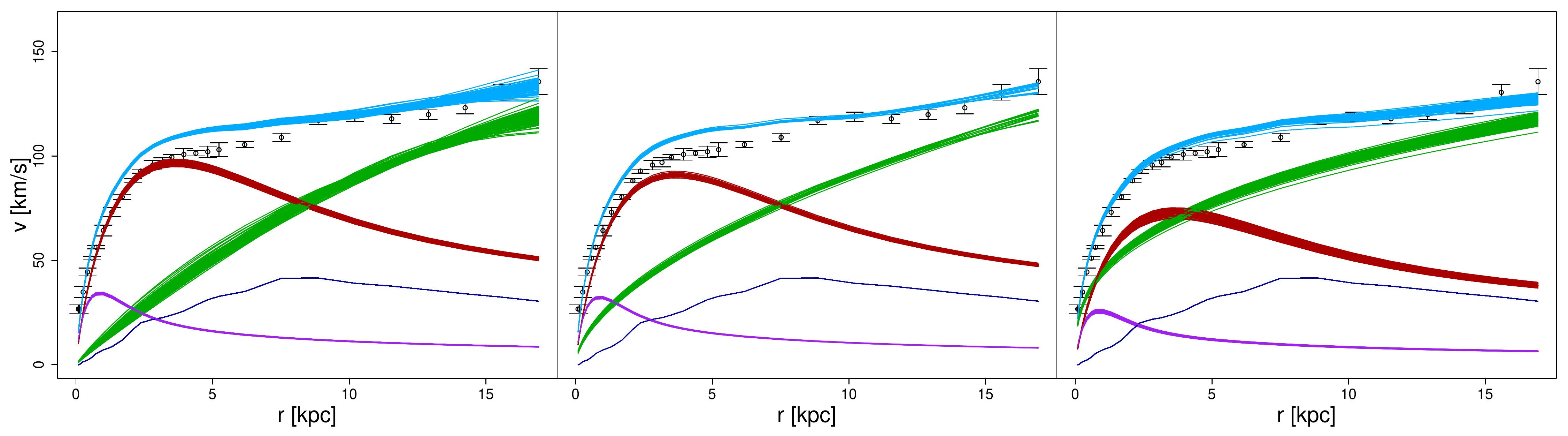}
  \caption{Overlay of representative samples of rotation curves from Figure \ref{gammainml} illustrating the degeneracy between halo slope and stellar mass-to-light ratio. From left to right the panels show models with $0.1<\gamma_{\rm in}<0.2$, $0.8<\gamma_{\rm in}<0.9$ and $1.3<\gamma_{\rm in}<1.4$, respectively. Key is as in Figure \ref{curve1}.}
  \label{bincurves}
\end{figure*}

We further illustrate the degeneracy by binning the MCMC models along the $\gamma_{\rm in}$ axis and showing a sample of rotation curves from each bin. This is presented in Figure \ref{bincurves}, which clearly shows the degeneracy between the disk and halo contributions. The right-hand panel shows the case of models near the peak of the probability distribution, and the left and middle panels show models from the tail towards shallower profiles. These are substantially disfavoured relative to the peak. 

\subsection{Impact of Rotation Curve Features}

The rotation curve presented in \cite{Corbelli:2000jg} and \cite{Corbelli:2003kn} shows an apparent feature in the outermost part of the rotation curve where the rotation velocity begins to increase, after having levelled off (see Figure \ref{curve1}). We now explore the extent to which these two data points influence our result on the distribution of $\gamma_{\rm in}$ by re-running our MCMC chains without the two outmost data points. This is required as we wish to show that our result is independent of the inclusion or not of these two data points.

The feature may be modelled by a flat dark matter density profile extending throughout the radial range, coupled with the maximal baryonic component to model the shape of the rotation curve at small $r$. We confirm this by calculating $r_1$, the radius at which the log slope of the dark matter halo reaches -1. We find that this value is high (on the order of the radial extent of the data) for high values of $\Upsilon_{3.6}$ (which from Fig.~\ref{gammainml} correspond to $\gamma_{\rm in} \approx 0$).

If these outermost data points represented a genuine feature of the density profile and the disk were not maximal, it would require an anomalous increase in the dark matter density at this point as the baryonic component is marginal here. As there is no obvious mechanism to form such a shell of dark matter, we cannot take this model to be correct at large $r$. 

We consider an artefact of the tilted ring method used to generate this rotation curve to be more a likely explanation of this rotation curve feature. In \cite{Corbelli:2000jg}, an initial set of radial bins in this rotation curve are generated by fitting a parameterised ring to the HI velocity field of the galaxy, under the assumption of entirely circular motion, and then additional radial bins are calculated from interpolating between the neighbouring rings. If the assumptions of the model do not hold e.g. in the presence of significant radial motion, then the parameters of a particular ring may be invalid. An underestimate of inclination would lead to an overestimation in rotation velocity, and due to the interpolation, a single incorrect inclination can account for the apparent feature seen in the last two radial bins. Considered without the final two radial bins, it is not clear the feature exists at all.

In \cite{Corbelli:2014wf} there are data that cover a greater radial extent than earlier papers. However, as can be seen in Figure 5 of that paper, data further out than the outmost limit of the rotation curve used here (from ~15kpc) have large errors and can be clearly approximated by a flat rotation curve. This confirms the above assessment, and means that there would be little to be gained for our specific goal of constraining $\gamma_{\rm in}$ by using the more extended rotation curve.

\subsection{Alternate mass-to-light priors}

\begin{figure}
  \includegraphics[width=\figwidth\linewidth]{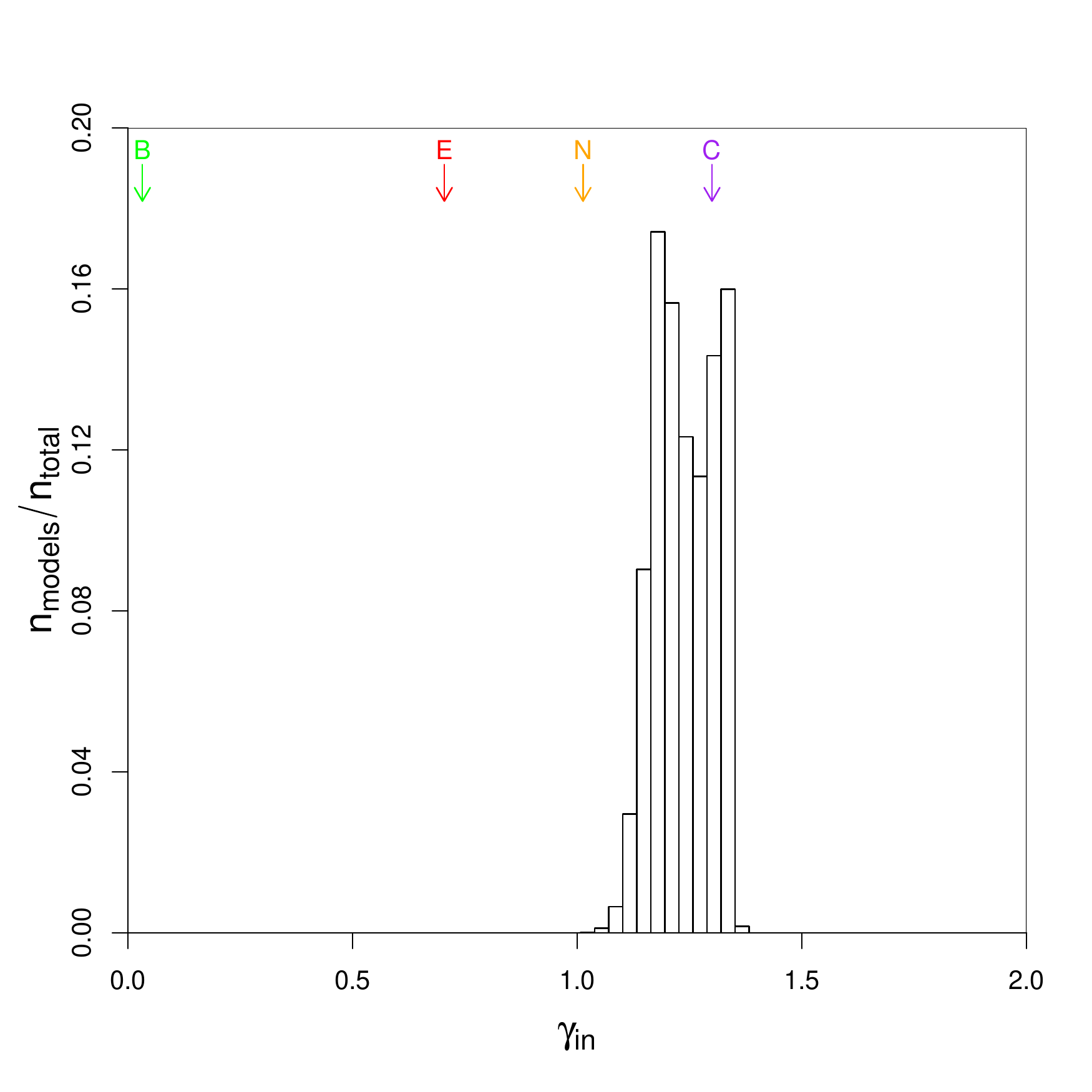}
  \includegraphics[width=\figwidth\linewidth]{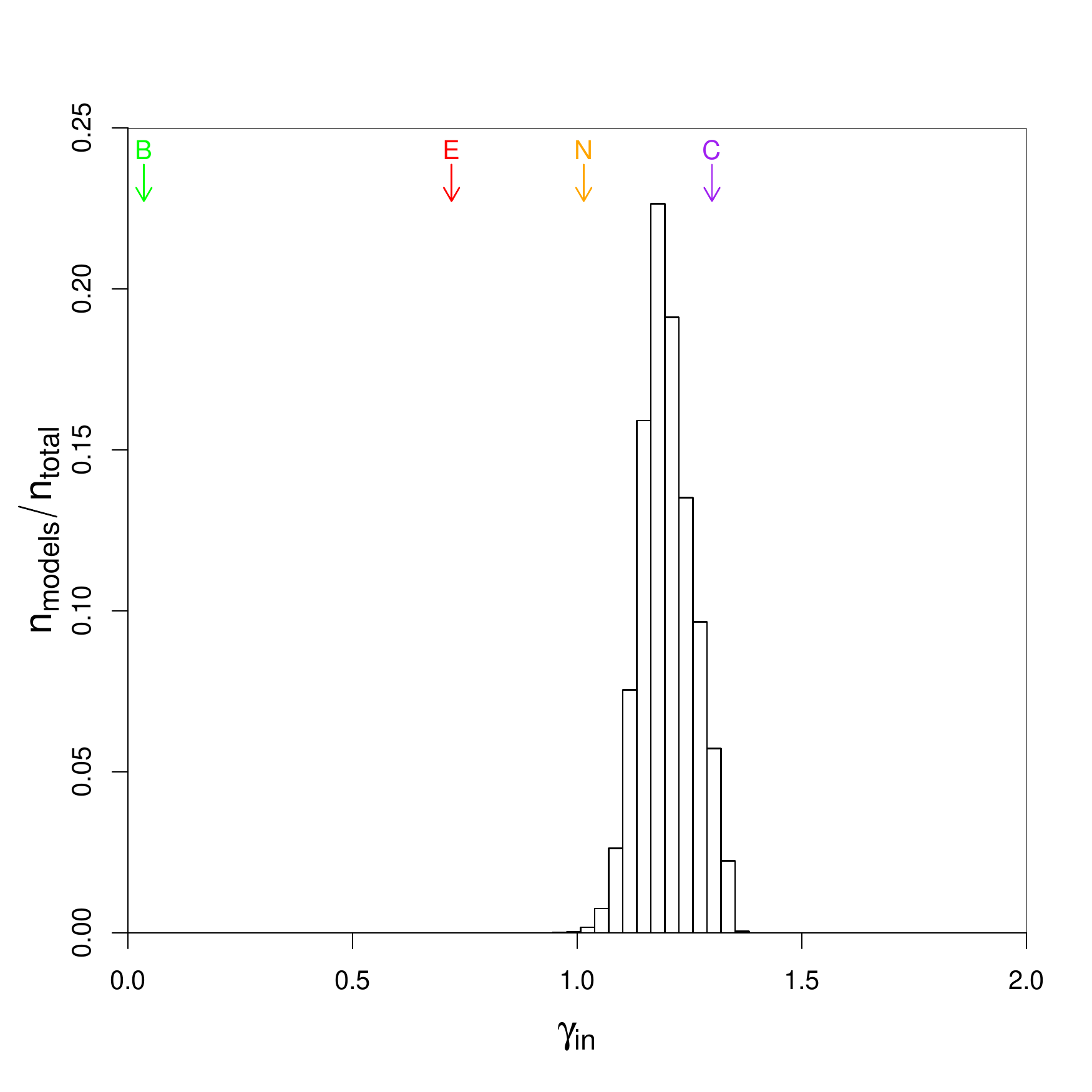}
  \caption{Histogram of values of $\gamma_{\rm in}$ in case B1 where $\Upsilon_{3.6}$ is fixed at 1.25 \citep[i.e. using the mass-to-light value from ][]{Seigar:2011gm}. Vertical axis shows number of models, normalised to give the histogram a total area of 1. {\bf Top} is the case for the full rotation curve and {\bf bottom} is the case where the last two data points are excluded. Arrows show log slopes for maximum likelihood fits of four individual profiles: green is the Burkert profile, red is the Einasto profile, orange is the NFW profile, and purple is the single power-law found in \cite{Corbelli:2000jg}. Note that the value calculated in \cite{Corbelli:2014wf} corresponds closely to the value for the NFW profile shown here.}
  \label{fixednormgammainhist}
\end{figure}

Figure \ref{fixednormgammainhist} shows that for run B1 (where $\Upsilon_{3.6}$ is taken from \cite{Seigar:2011gm}), the distribution of $\gamma_{\rm in}$ favours a cusped density profile, steeper than the best fitting NFW profile. This does not imply that an NFW halo does not fit the rotation curve, merely that other regions of parameter space are favoured. The distribution is bimodal, but removing the last two data points removes the second peak (lower panel). This peak is then purely dependent on a feature which may be an artefact. 

\begin{figure}
  \includegraphics[width=\linewidth]{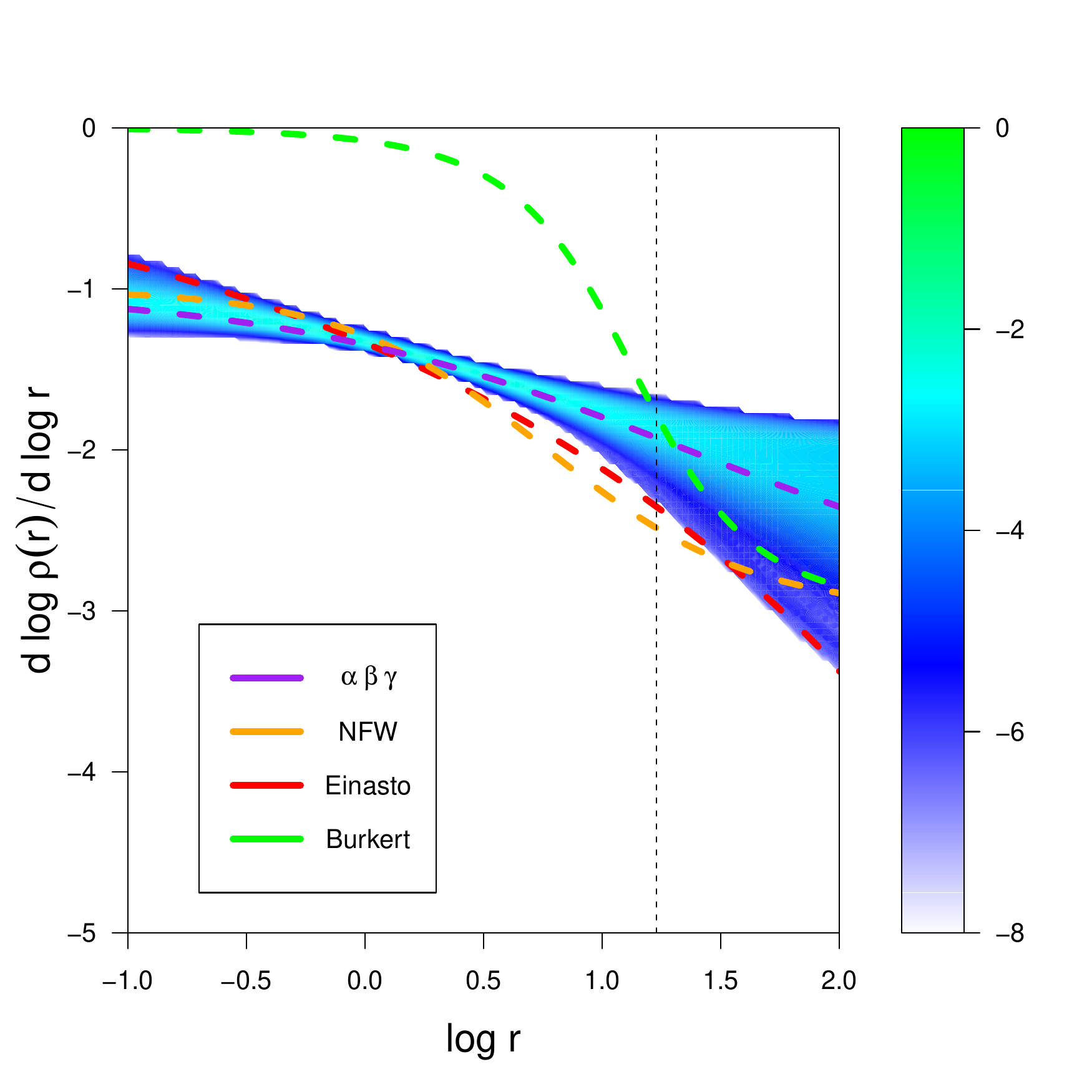}
  \caption{Density plot of all the models produced by C1, in $d {\rm log} \rho/d {\rm log} r$ space. Key as in Figure \ref{overplot}.}
  \label{lowoverplot}
\end{figure}

In runs C1 and C2, we found a log slope compatible with the NFW profile. However, this was in a region that is disfavoured by the run with a free value of $\Upsilon_{3.6}$. In Figure \ref{lowoverplot} we show the equivalent result to Figure \ref{overplot}, which demonstrates that in this case a single power law is not favoured. The smaller stellar contribution to the kinematics requires the shape of the rotation curve to be primarily modelled by the dark matter halo, and the NFW profile is unable to do this for the entire radial range. The inner part (log $r < 0.5$) would require a different concentration parameter $c_{\rm vir}$ than the outer part (log $r > 0.5$) and in the fitting statistic $\chi^2$ is weighted towards the inner part of the galaxy as there are more data points there.

\begin{figure}
  \includegraphics[width=\figwidth\linewidth]{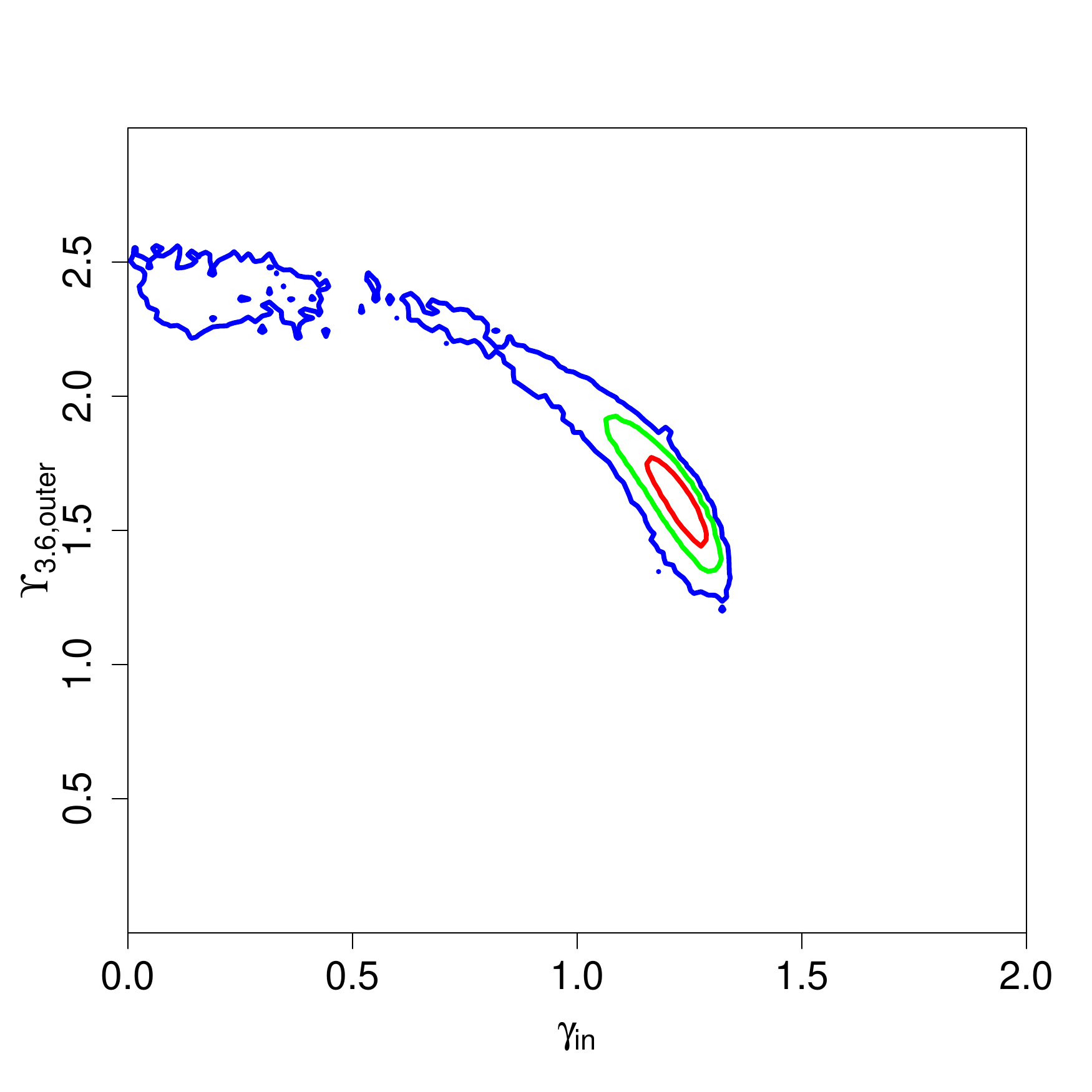}
  \includegraphics[width=\figwidth\linewidth]{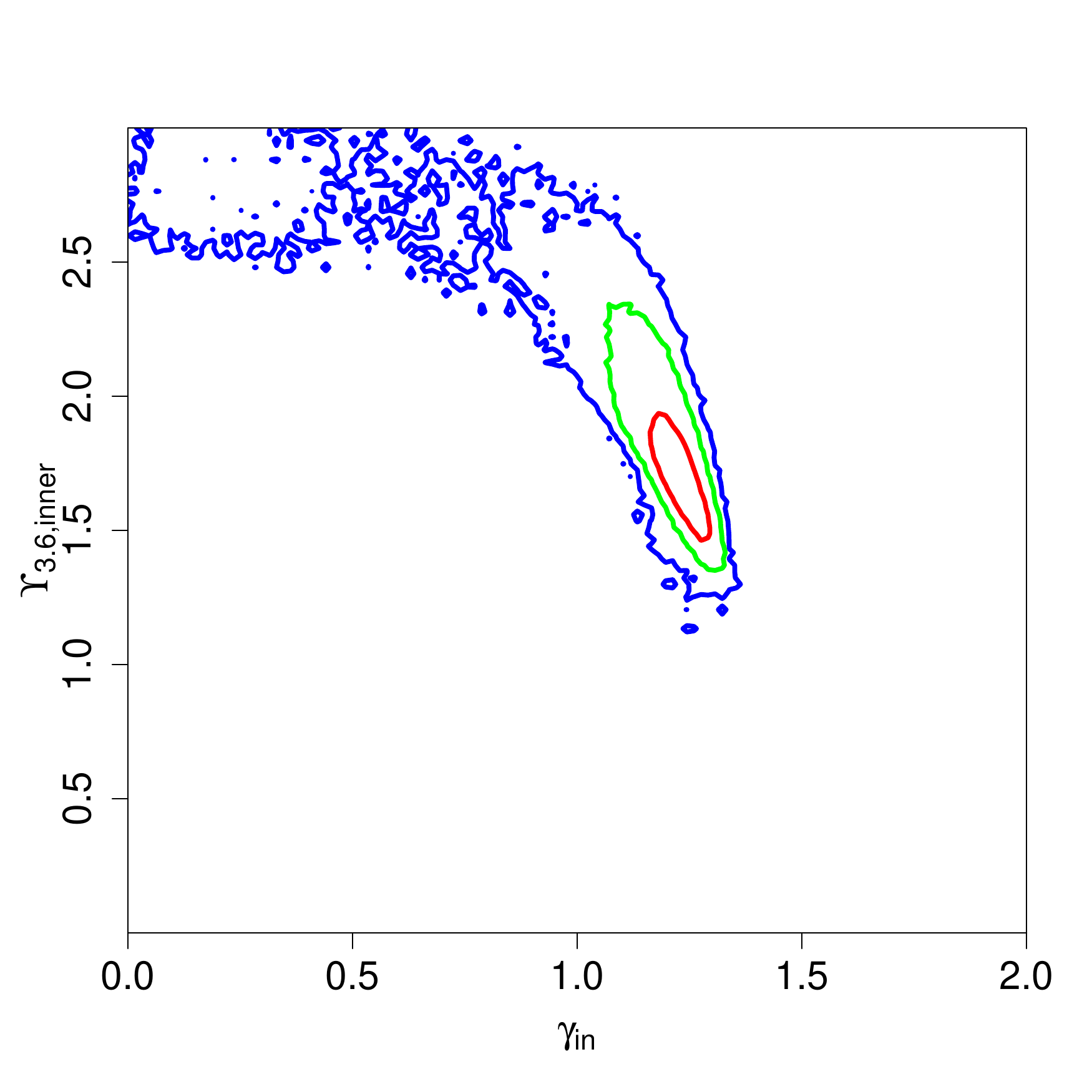}
  \caption{For run D1. Correlation between $\gamma_{\rm in}$ and {\bf(top)} mass-to-light ratio $\Upsilon_{3.6}$ for the outer stellar component and {\bf (bottom)} $\Upsilon_{3.6}$ for the inner stellar component. There is a prior constraint that the inner mass-to-light ratio be higher.}
  \label{corefig}
\end{figure}

Runs D1 and D2, with an additional free parameter for the $\Upsilon_{3.6}$ value of the inner stellar component, produced a similar degeneracy to the A runs. Figure \ref{corefig} shows the relation between $\gamma_{\rm in}$ and both values of $\Upsilon_{3.6}$. There is a weak, secondary peak of models featuring a maximal inner component and a flat ($\gamma_{\rm in}<0.5$) inner slope, but the main peak in $\Upsilon_{\rm 3.6, outer}$ versus $\gamma_{\rm in}$ is unaffected. We repeated this test, imposing a prior that the value of $\Upsilon_{3.6}$ for the inner component always be greater than that for the outer component, and found that the result shown in Figure \ref{corefig} remained unchanged. It should be noted that the smooth variation of these two components effectively parameterises a mass-to-light gradient across the stellar disk (which we elected not to model, see Section{massmodels}) and demonstrates that its inclusion would not impact the result. In general, we find that cusped halo models exhibit smaller gradients than cored models. The distribution of $\Upsilon_{\rm 3.6, inner}$ is unsurprisingly wide, due to its dependence on a small number of the innermost data points. 

\section{Discussion}
\label{discusssec}

\subsection{Comparison with Previous Work}

We have found a result more in agreement with the value of $\gamma_{\rm in}=1.3$ implied by \cite{Corbelli:2000jg} than the assertion by \cite{Seigar:2011gm} that M33 is best described by an NFW halo. Whilst the NFW halo does fit the rotation curve decomposition with $\chi^2_{\rm red} = 1.18$, there are many other haloes (mostly steeper) that also fit the same data equally well. 

We noted in HW13 that with multi-parameter models, $\chi^2_{\rm red}$ is not reliable across the entire parameter space because it is calculated assuming that the degrees of freedom are constant across the parameter space. This  cannot be assumed to be the case. For instance, if a model includes a high stellar mass, and a reduced contribution of the dark matter halo to the rotation curve at small $r$, then any shape parameters of the halo are going to become less relevant to the quality of the fit. This was described, in an extreme case, in HW13 for the case of constructed high surface brightness rotation curves where the dark matter contribution to the rotation speed was smaller than the error bars. Even in less extreme cases, the parameter $\beta$ is often not fully utilised, if the scale radius of the halo is large enough that $\beta$ does not become the dominant shape parameter within the data range. Model comparison on the basis of 
$\chi^2_{\rm red}$ alone is thus not necessarily meaningful.

The probability of adding a model, a point in parameter space, to one of our MCMC chains is not based on the absolute value of its $\chi^2$ but on the gradient of the goodness of fit, i.e. the relative goodness of fit compared to some other point the chain may arrive from. The final result is essentially an integral of this value over all possible starting points - but with a substantial weighting towards nearby points due to the Gaussian shape of the model selection function in the Metropolis-Hasting algorithm. This means that the MCMC result does not rely on the goodness of fit being a globally correct representation of likelihood. Given that in a tilted ring model, errors are computed from azimuthal variations in the inferred circular velocity, it is not immediately clear they satisfy the requirements of being independent, Gaussian errors as assumed when $\chi^2$ is used to calculate a statistically robust likelihood. An analysis (such as ours - see HW13) that does not rely on the assumption of Gaussian errors is preferable. 

In HW13 we showed that the MCMC method was able to recover the correct halo model from synthetic data with artificially inflated error bars more tightly than would be naively expected. We attribute this not only to the method being less dependent on the actual values of the goodness of fit, but also to the physically reasonable prior assumption of a smooth dark matter density profile. Smoothness is inherent in the $\alpha-\beta-\gamma$ profile for reasonable values of $\alpha$, but for an even freer prior, smoothness would have to be imposed separately. A non-parametric halo, with a log slope for each data point in the rotation curve, would have to impose a constraint on these values such that together they form a physically realistic density profile.

In \cite{Seigar:2011gm} it is claimed that fitting of the NFW profile to rotation curve bins outside $r=7{\rm kpc}$ is evidence that the this profile "best represents these data". Only the prior assumption of an NFW profile makes the fit to the outer rotation curve relevant to the determination of the inner density profile. Without assuming a strong link between inner and outer data points as a prior, we find that a steeper inner density profile is favoured when using the $\Upsilon_{3.6}$ value used in that paper.

The fact that the NFW profile is able to fit the rotation velocities in previous work does not itself convey the properties of the NFW profile (i.e. a log slope of -1 at $r=0$) upon the galaxy, given that we have shown there are many profiles with differing properties that also provide good fits. Thus, it cannot be concluded that a low reduced $\chi^2$ value for an NFW fit gives a high posterior probability for specific analytic properties of the NFW profile, e.g. the central $\rho \propto r^{-1}$ cusp. Allowing more freedom in the profile, and fully exploring the parameter space with MCMC, resolves these issues and provides a more robust description of the dark matter density profile across the entire radial range of the rotation curve data.

The second claim of \cite{Seigar:2011gm} is that the NFW concentration parameter $c_{\rm vir}$ is related to the spiral arm pitch angle $P$. Taking $c_{\rm vir}$ purely as a density profile parameter without any implications for inner profile slope, correlations between it and other physical parameters are not necessarily in conflict with our conclusions.

\subsection{$\Upsilon_{3.6}-\gamma_{\rm in}$ Degeneracy}

It is clear from our result that there is a substantial degeneracy between the baryonic mass-to-light ratio $\Upsilon_{3.6}$ and the inner log slope $\gamma_{\rm in}$. The scale length of the stellar disk is approximately equal to the radius at which the total rotation curve transitions from rising to flat, so the most obvious cause of the degeneracy is the degree to which the rising part of the stellar contribution to the circular velocity is used to model the rising part of the overall rotation curve. To investigate this, we assume that the degeneracy of the fit near the peak of the disk rotation curve approximates that of the entire rotation curve, and that the inner dark matter halo can be taken to be a single power law (which in our results it can; over 98\% of models in A1 have a scale radius outside the data range). We can then represent the degeneracy using

\begin{equation}
v_{\rm max, star}^2 + v_{\rm DM} (R_{\rm max})^2 = {\rm constant}
\end{equation} 

where $v_{\rm max,star}$ is the maximum velocity contribution of the stellar component and $v_{\rm DM} (R_{\rm max})$ is the velocity contribution of the dark matter halo at that radius. Given that  $v_{\rm max,star}^2 \propto \Upsilon_{3.6}$ and $v_{\rm DM} (R_{\rm max})^2 \propto (R_{\rm max}/r_{\rm s})^{2-\gamma_{\rm in}}$ (as $\gamma_{\rm in}$ is assumed to be representative of the log slope at this radius, due to our assumption of a single power law) we can write

\begin{equation}
\Upsilon_{3.6} = a -  b c^{2-\gamma_{\rm in}}
\label{degeneq}
\end{equation}

where $a$, $b$, and $c$ are constants. This is not meant to imply that the galaxy is well fitted by a single power law; this is merely meant to approximate the dark matter halo density profile well in the region which is most sharply affected by the degeneracy. The curve in $\gamma_{\rm in}-\Upsilon_{\rm 3.6}$ space described here is not a physical description, nor a prediction, but is useful as a parameterised, quantitative summary of the degeneracy determined by the MCMC method. 

\begin{figure}
  \includegraphics[width=\linewidth]{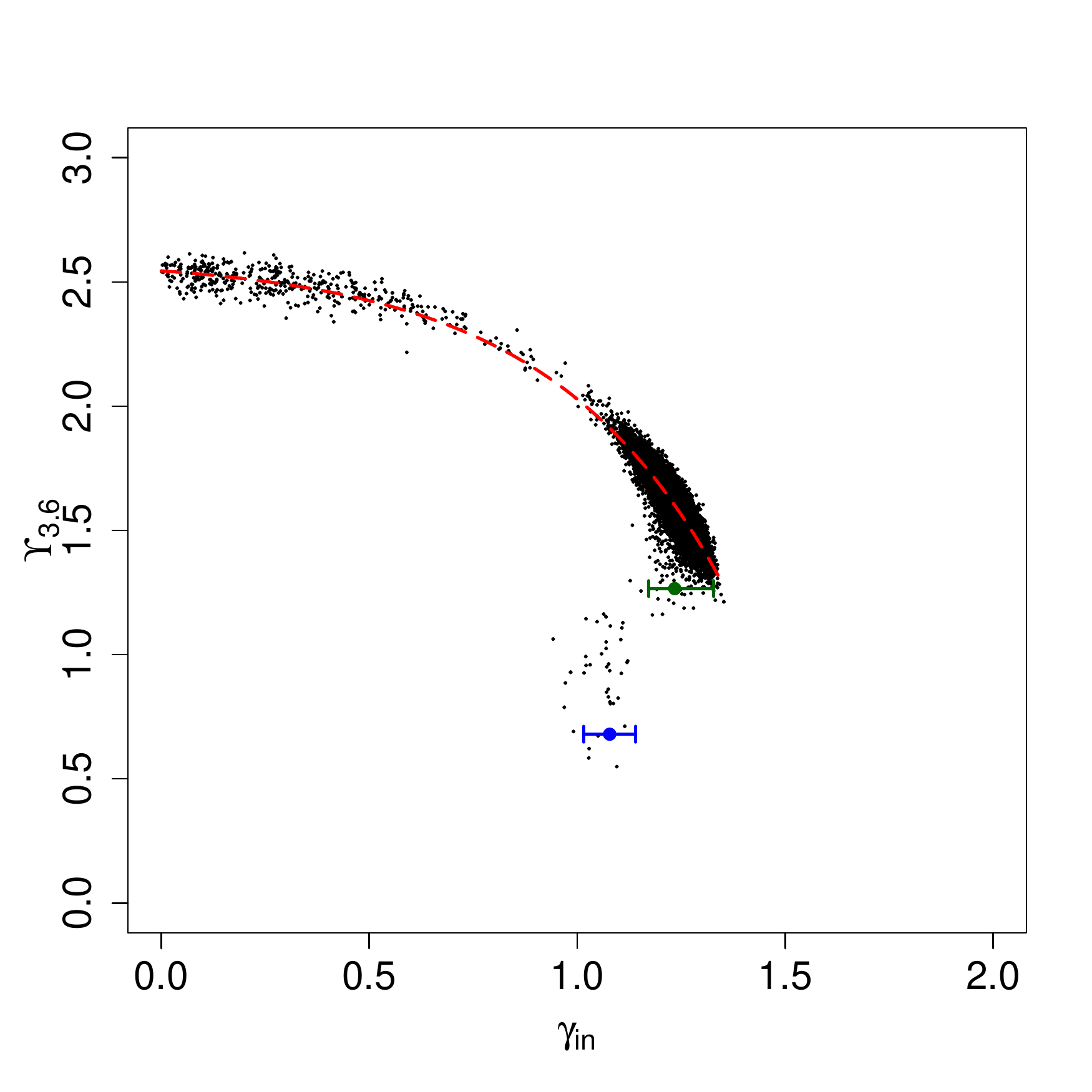}
  \caption{Fit to the degeneracy between log slope at the inner most bin ($\gamma_{\rm in}$) and mass-to-light ratio in the Spitzer 3.6$\mu m$ band ($\Upsilon_{3.6}$). Points are 34846 models chosen from the chain, after burn-in, with a probability $10^{-5}$. The line shows the relationship $\Upsilon_{3.6} = 2.6 -  6.3 \times 0.089^{2-\gamma_{\rm in}}$, fitted to 2000 bins, and should not be taken as valid outside the range of the models shown. Green and blue points are constraints on $\gamma_{\rm in}$ found by models B1 and C1 respectively. See text for discussion.}
  \label{degenfit}
\end{figure}

We binned $\gamma_{\rm in}$ values of a subset of models from A1 (34846 models chosen from all chains with a probability $10^{-5}$, ignoring the first 10,000 models for burn-in) to produce a set of 2000 bins with a uniform number of points. We fit the relationship (\ref{degeneq}) to these data and find $a=2.6$, $b=6.3$, and $c=0.089$ as shown in Figure \ref{degenfit}. The baryonic mass-to-light ratio in the maximal disk case is $\Upsilon_{\rm 3.6, max}=a-bc^2$ and the halo slope in the no disk case is $\gamma_{\rm in, no} = 2 - ({\rm log} \,a - {\rm log} \,b)/{\rm log} \,c$.

In runs D1 and D2, we tested whether this degeneracy would be changed by having independent $\Upsilon_{3.6}$ values for both stellar components. The scale radius of the inner component is too short to meaningfully contribute to the circular velocity at the radius at which we model the degeneracy between $\gamma_{\rm in}$ and $\Upsilon_{\rm 3.6, outer}$ above. 

\subsection{Alternate Gas Mass Models}

We now consider whether it is possible that the apparently large stellar mass in our cored models could be accounted for by molecular gas. The gas density profile we obtained from \cite{Corbelli:2003kn} included the molecular gas fraction, calculated from CO emission, along with the atomic gas contribution to the rotation curve. The factor used to calculate total molecular gas mass from CO emission, $X_{\rm CO}$, is estimated based on observations of the Milky Way \citep{Wilson:1995kc} and may not be correct for M33. In the mass modelling we use, the stellar disk has a mass $3.8\times10^9 M_\odot$ and the molecular gas disk component has a mass $3\times10^8 M_\odot$, so this would represent an increase in $\Upsilon_{3.6}$ of $\sim0.1$. (Note that, following ~\cite{deblok2008}, we have assumed that the scale length of the molecular gas is that same as that of the stars). To account for the difference between the modelled $\Upsilon_{3.6}=0.67$ and the $\Upsilon_{3.6} = 2.5$ required for a flat halo would require $X_{\rm CO}$ to be 37 times larger. In \cite{Dame:2001bg} the $1\sigma$ relative error for this ratio was found to be less than 0.17 for nearby clouds in the Milky Way and thus a factor 37 increase seems unlikely.

\subsection{Comparison with cosmological simulations}

Given that we have confirmed earlier claims that the halo profile of M33 is steeper than in other galaxies of similar luminosity, it is worth asking whether there is a natural explanation for this difference. In their models of the Local Group \cite{Bekki:2008ek} found that M33 encountered M31 with a periapsis of $\sim100{\rm kpc}$ at 4-8Gyr before present. Ram pressure stripping of M33 by an outflow from M31 could also have had an impact on the progression of feedback in M33 \citep[e.g. ][]{Nayakshin:2013im}. The prompt removal of low density gas by stripping would cause the dark matter halo to relax into a shallow, less concentrated state \citep[as modelled in][and HW14]{Gnedin:2002bu, Read:2005hx, Governato:2010ed}, so it might be reasonably assumed that such an event would lead to a cored dark matter halo, but this is not necessarily the case, as we now discuss. 

It is reasonable to assume the the process of contraction would mean that after initial baryon infall, but before feedback begins, the dark matter halo would have an inner log slope steeper than $\gamma_{\rm in}=1$. Feedback models such as \cite{Read:2005hx}, \cite{Governato:2010ed}, \cite{Parry:2011fd} and \cite{Ogiya:2012tq} require multiple outflow and inflow events to account for the transition from such steep initial haloes to their  flatter inner haloes at later times. If this process were interrupted early on, it could prevent a sufficiently large amount of feedback that, even though the event itself would flatten the halo slightly, it would still retain an inner halo that is steep relative to those of similar galaxies. 

\section{Conclusion}

We have modelled the rotation curve of M33 using the MCMC-based approach we presented in HW13. We have quantified and understood the degeneracy between baryonic mass-to-light ratio $\Upsilon_{3.6}$ and the log slope of the dark matter halo at the inner bin $\gamma_{\rm in}$. We cannot resolve the conflict between observations of similar galaxies (\cite{KuziodeNaray:2006ez} and \cite{deNaray:2008iz}) and the MCMC analysis of the M33 rotation curve without assuming $\Upsilon_{3.6}>2$, which is difficult to reconcile with stellar population modelling. We find that with a lower fixed $\Upsilon_{3.6}$ = 0.67, an NFW halo is compatible with the data, but that this part of parameter space is not strongly favoured when we relax the constraint on $\Upsilon_{3.6}$. We strongly exclude the combination of $\Upsilon_{3.6}<2$ and a halo profile inner log slope $\gamma_{\rm in}<0.9$, for a comprehensive range of assumptions. 

The constraints we find on $\Upsilon_{3.6}$ and $\gamma_{\rm in}$ admit at least the four following scenarios:

\begin{enumerate}
\item there is a great deal more mass in the disk of M33 than is accounted for by standard modelling of stellar populations and molecular gas clouds.

\item the halo of M33 deviates significantly from spherical symmetry, being flattened at small disk radius and less so in the outer part of the galaxy. 

\item feedback cannot produce a core in a galaxy with the stellar mass of M33. \cite{DiCintio:2013em} make this point but their conclusion depends on the specific feedback physics used in that paper, and only accounts for supernova and early stellar feedback. It also predicts a shallower ($\gamma_{\rm in} \simeq 0.75$) inner density profile for a galaxy with the range of stellar masses we calculate.

\item the dark matter halo has a much steeper inner profile than would be expected from hydrodynamical simulations of galaxy formation \citep[e.g.][]{Governato:2010ed,Maccio:2011ge}. This could occur if M33 were dominated by the process of contraction. The above simulations show both contraction of the halo steepening the inner profile, and feedback flattening it. In the absence of any obvious source of significant additional disk mass, and assuming no fundamental error in the view of baryon-dark matter interaction in galaxy formation, we propose that the history of the dark matter halo in M33 is dominated by contraction. Ram pressure stripping by M31 before feedback flattening the halo is a possible physical mechanism by which this could have happened. 

\end{enumerate}

The first scenario is inconsistent with the stellar mass modelling of \cite{Corbelli:2014wf}. We will investigate the remaining three possibilities by applying our modelling scheme to cosmological simulations in a future work. 

\section{Acknowledgements}

This work is based [in part] on observations made with the Spitzer Space Telescope, which is operated by the Jet Propulsion Laboratory, California Institute of Technology under a contract with NASA. The CosmoMC code was written by Anthony Lewis~\citep{lewis2002}. We thank Edvige Corbelli for supplying the initial rotation curve decompositions. We acknowledge Wyn Evans for valuable discussions during a visit by MIW to the Aspen Center for Theoretical Physics, and Simon White for bringing work on M33 to our attention. We are grateful to the anonymous referee for detailed comments which improved the final version of the paper, and to Carlos Frenk for useful comments. MIW acknowledges the Royal Society for support through a University Research Fellowship. PRH acknowledges STFC for financial support. This research used the Complexity HPC cluster at Leicester which is part of the DiRAC2 national facility, jointly funded by STFC and the Large Facilities Capital Fund of BIS.

\bibliographystyle{apj}
\bibliography{m33}

\end{document}